\documentclass[twocolumn,showpacs,pra,superscriptaddress]{revtex4}
\usepackage{dcolumn}
\usepackage{graphicx}
\usepackage{bbm}

\newcommand{\be}{\begin{equation}}
\newcommand{\ee}{\end{equation}}
\newcommand{\beq}{\begin{eqnarray}}
\newcommand{\eeq}{\end{eqnarray}}

\newcommand{\hide}[1]{}

\newcommand{\fig}[1]{Fig.\,\ref{#1}}

\newcommand{\ket}[1]{\left| #1 \right\rangle}

\begin{document}

\title{Phase gate and readout with an atom/molecule hybrid platform}

\author{Elena Kuznetsova}
\affiliation{Department of Physics, University of Connecticut,
Storrs, CT 06269}
\affiliation{ITAMP, Harvard-Smithsonian Center
for Astrophysics, Cambridge, MA 02138} 
\author{Marko Gacesa}
\affiliation{Department of Physics, University of Connecticut,
Storrs, CT 06269}
\author{Susanne F. Yelin}
\affiliation{Department of Physics, University of Connecticut,
Storrs, CT 06269} 
\affiliation{ITAMP, Harvard-Smithsonian Center
for Astrophysics, Cambridge, MA 02138}
\author{Robin C\^ot\'e}
\affiliation{Department of Physics, University of Connecticut,
Storrs, CT 06269}
\date{\today}

\begin{abstract}

We suggest a combined atomic/molecular system for quantum computation, which takes advantage of highly developed 
techniques to control atoms and recent experimental progress in manipulation of ultracold molecules.
We show that two atoms of different species in a given site, {\it e.g.}, in an optical lattice, could be used for qubit encoding, initialization and 
readout, with one atom carrying the qubit, the other enabling a gate. In particular, we describe how a two-qubit phase gate can be 
realized by transferring a pair of atoms into the ground rovibrational state of a polar molecule with a large dipole moment, and allowing two 
molecules to interact via their dipole-dipole interaction. We also discuss how the reverse process of coherently transferring a molecule into a pair of 
atoms could be used as a readout tool for molecular quantum computers.

\end{abstract}

\maketitle

In recent years, hybrid systems taking advantage of specific characteristics of given platforms for quantum information processing \cite{Hybrid} 
have generated increasing interest. Neutral-atom based quantum computing \cite{Neut-atoms-QC} offers long-lived atomic states ({\it e.g.} hyperfine) 
to encode qubits and well-mastered techniques to manipulate atomic transitions. The usually weak coupling between atoms, however, would lead to long 
two-qubit gate operations, and two main approaches are studied to overcome this issue: using intermediate Rydberg states \cite{Rydberg-phase-gate}, 
or on-site collisions \cite{Collisional-phase-gate,Tommaso}. 
The large Rydberg-Rydberg interactions allow for rapid gate operations, but suffer from the Rydberg atoms' strong interaction with the environment 
and radiative decay, while the need to coherently move atoms in an optical lattice to bring them to one site slows the collisional gates. A different platform, 
polar molecules \cite{DeMille}, offers long-lived states and direct strong dipole-dipole interaction, potentially resulting in fast robust two-qubit gates. 
However, readout and initialization of molecular qubits still poses a problem.

In this Letter, we report on a basic atom-molecule transfer procedure that allows, on one hand, to realize efficient two-qubit operations for 
neutral atoms and, on the other hand, to initialize and readout qubits encoded in states of polar molecules.  
Specifically, we envision an optical lattice with two 
atomic species per site in a Mott insulating phase: a qubit is encoded in hyperfine states of one atom, while the second one serves as an ``enabler" 
to allow for conditional atom/molecule conversion. 
Qubit initialization, storage, readout, and one-qubit operations are implemented using well-developed techniques of atomic states manipulation.
To execute a two-qubit gate, such as a phase gate, a pair of atoms in one site is converted into a stable molecule with a large dipole moment.  
It can then interact with 
a molecule formed at another site via long-range strong dipole-dipole interaction, which effectively ``switches" on the interaction between selected 
qubits only when needed \cite{PRAs}. 
We note that an optical lattice with two different atoms per site has been experimentally 
demonstrated along with state-selective conversion of 
the atomic pair into a weakly bound Feshbach molecule \cite{KRb-lattice}, pointing to the feasibility of the proposed scheme 
with current experimental techniques. 
An alternative application of this scheme is the readout of molecular qubits. 
Namely, qubits can be stored in long-lived molecular states as in Refs. \cite{DeMille,PRAs}, and conditionally converted into atoms for readout. This is a non-destructive 
alternative to selective state ionization: the readout atoms can be converted back into molecules which can be used again.

The characteristics needed for our hybrid platform are (i) long-lived atomic states, (ii) collisions between two atoms in a lattice 
site be elastic, (iii) possibility to controllably and reversibly convert atoms into stable molecules, 
and (iv) molecules with large dipole moments.  Many systems exhibit these properties, specifically, alkali atoms have hyperfine states possessing 
long coherence times, and molecules formed from alkali atoms of different species typically have a large permanent electric dipole 
moment in the ground state. Weakly bound molecules can be formed from ultracold atoms by photo or magnetoassociation \cite{Feshbach}, 
and coherently transferred to deeply bound states by optical pulses using STIRAP (Stimulated Raman Adiabatic Passage) \cite{Ni} or a Raman $\pi$ pulse. 
In the following we illustrate the scheme using an atomic 
pair of $^{87}$Rb and $^{7}$Li.

\begin{figure}
\center{
\includegraphics[width=9.cm]{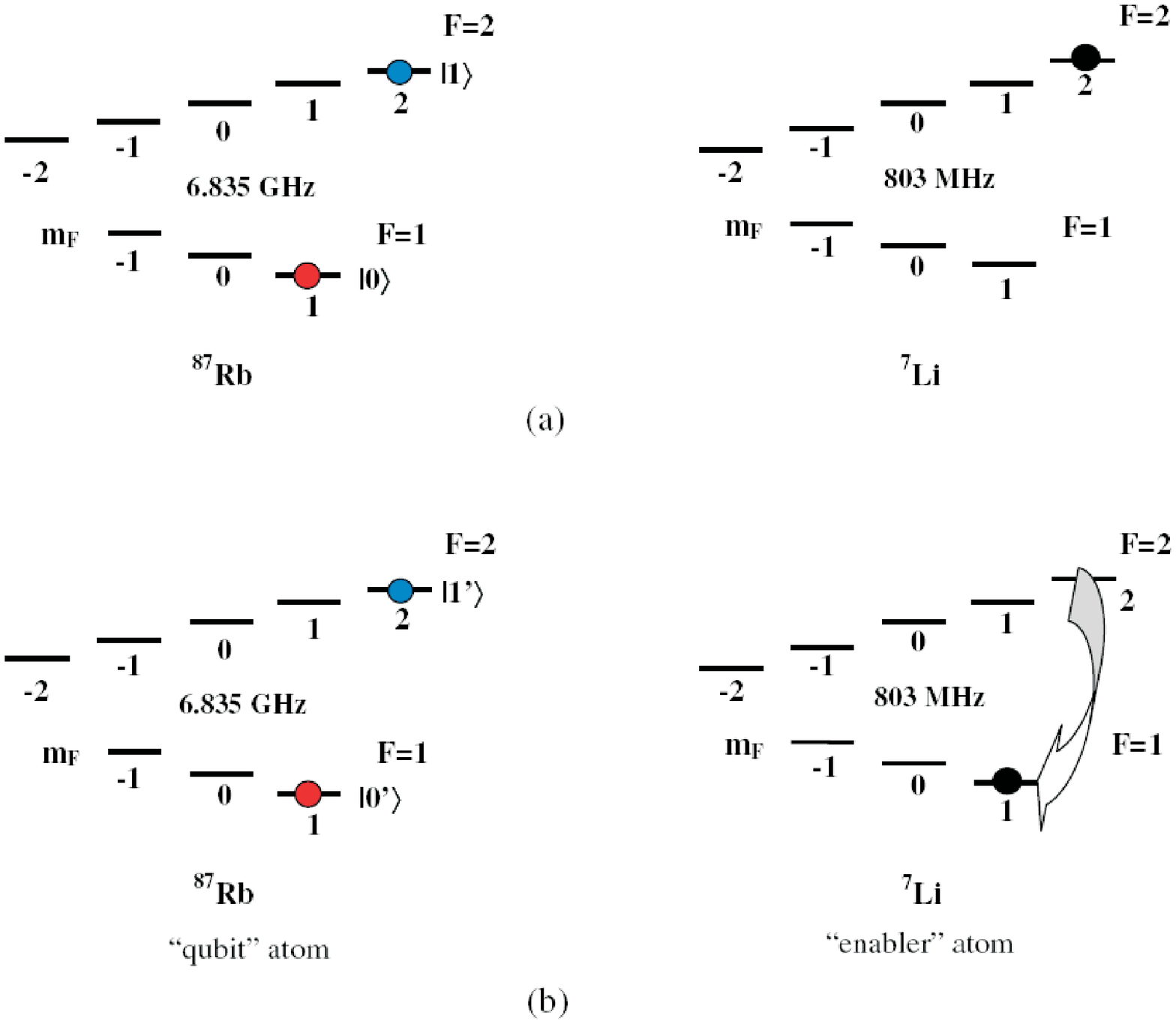}
\caption{\label{fig:Rb} (a) During storage and one-qubit operations qubit states are encoded into hyperfine 
sublevels $\ket{f,m}=\ket{2,2}$ and $\ket{1,1}$ of $^{87}$Rb, while $^{7}$Li is kept in an ``inert" state $\ket{2,2}$; 
the qubit states are products $\ket{0}=\ket{1,1}^{\rm Rb}\otimes \ket{2,2}^{\rm Li}$ and $\ket{1}=\ket{2,2}^{\rm Rb}\otimes \ket{2,2}^{\rm Li}$. 
(b) To perform a two-qubit phase gate the ``enabler" $^{7}$Li atom can be transferred into the $\ket{1,1}$ state, the system is 
then in a superposition of ``enabled" qubits $\ket{0'}=\ket{1,1}^{\rm Rb}\otimes \ket{1,1}^{\rm Li}$ and $\ket{1'}=\ket{2,2}^{\rm Rb}\otimes \ket{1,1}^{\rm Li}$. 
A Feshbach resonance, which enhances the efficiency of atom-molecule conversion, takes place for the $\ket{1,1}^{\rm Rb}+\ket{1,1}^{\rm Li}$ collision channel.}
}
\end{figure}

{\it{Qubit states.}}
A qubit can be encoded into hyperfine sublevels $\ket{f,m}$ of one 
atom, while the state of the other atom is such that all scattering processes are elastic. In the Rb-Li system, $^{87}$Rb can provide 
qubit states, while $^{7}$Li can serve as the ``enabler". The $\ket{2,2}$ and $\ket{1,1}$ states of $^{87}$Rb can be used for qubit encoding, 
and $^{7}$Li can be kept in $\ket{2,2}$ during storage and one-qubit operations. 
In these states, Rb and Li collide elastically at ultralow temperatures, avoiding atomic loss and qubit decoherence. 
For the two-qubit gate the ``enabler" Li can be transferred into the $\ket{1,1}$ 
state, where a Feshbach resonance occurs at $B\sim 649$ G for the $\ket{1,1}^{\rm Rb}+\ket{1,1}^{\rm Li}$ collision channel. 
As we show below, this resonance can strongly enhance the efficiency 
of the atom-molecule conversion. To summarize, qubit states during storage and one-qubit operations and ``enabled" (primed) qubits during a two-qubit gate are:

\begin{eqnarray}        
\ket{0} \equiv \ket{1,1}^{\rm Rb}\otimes \ket{2,2}^{\rm Li},\; \ket{1} \equiv \ket{2,2}^{\rm Rb}\otimes \ket{2,2}^{\rm Li},\\
\ket{0'} \equiv \ket{1,1}^{\rm Rb}\otimes \ket{1,1}^{\rm Li},\; \ket{1'} \equiv \ket{2,2}^{\rm Rb} \otimes \ket{1,1}^{\rm Li}. 
\end{eqnarray}
 One-qubit rotations can be implemented using optical Raman pulses resonant with the qubit transition of Rb while not affecting the ``enabler" which is 
far-detuned (see \fig{fig:Rb}). 

\begin{figure}
\center{
\includegraphics[width=7.cm]{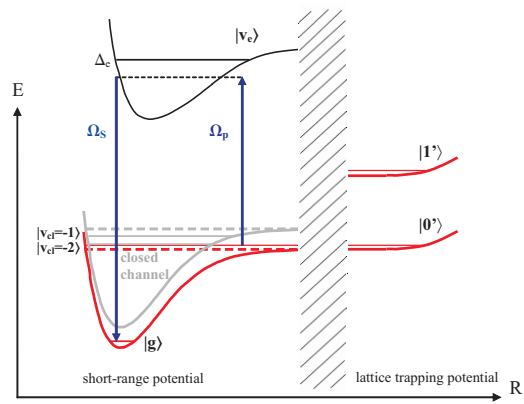}
\caption{\label{fig:STIRAP} Conversion of two atoms into a ground state molecule. Atoms in the lowest motional state of the collision channel 
corresponding to a qubit state $\ket{i}$ ($\ket{0}$ or $\ket{1}$) can be coherently transferred into a deeply bound molecule $\ket{g}$ by an off-resonant Raman $\pi$ pulse. 
Selectivity of transfer can be realized by setting pump ($\Omega_{p}$) and Stokes ($\Omega_{S}$) laser fields in a 
two-photon resonance with the $\ket{i}-\ket{g}$ transition. The dipole moment of the $\ket{i}-\ket{v_{e}}$ 
transition can be enhanced by a Feshbach resonance, {\it i.e.} by admixing a bound molecular 
state of a closed channel (shown in grey) to a scattering atomic state of the open collision channel (shown in red). 
As an illustration a second bound vibrational state $\ket{v_{cl}=-2}$ 
is assumed to be resonant with $\ket{0'}$.}
}
\end{figure} 

{\it Two-qubit gate.}
For a two-qubit gate we conditionally transfer a pair of atoms into a stable molecular state with a 
large permanent electric dipole moment. An external electric field orients the dipole moment in a lab frame and molecules in two different sites 
can interact via their dipole-dipole interaction. After a $\pi$ phase shift is accumulated, the molecules are converted back into atoms, realizing a 
phase gate. (Combining the phase gate with two $\pi/2$ rotations applied to a target qubit implements a CNOT gate.) 
Two unbound atoms can be coherently transferred into a molecular state by two-color photoassociation, {\it e.g.} using an optical off-resonant Raman $\pi$ pulse or by STIRAP, 
as shown in \fig{fig:STIRAP}. Conditional transfer can be, in principle, realized by selectively addressing qubit states by their frequency, {\it i.e.} 
by setting the laser fields 
in a two-photon resonance with the $\ket{1}-\ket{g}$ transition, where $\ket{g}$ is the ground rovibrational molecular state. In 
this case the $\ket{0}-\ket{g}$ transition will be far-detuned and no transfer 
will take place.

{\it Molecular qubit readout.}
A different application of the conditional atom-molecule conversion could be for readout of molecular qubits. Molecules possess long-lived 
rotational and spin sublevels in the ground electronic state to encode qubits. 
Specifically, molecules with a singlet $^{1}\Sigma^{+}$ ground state, which is typical for diatomics, have a zero electronic spin. 
The nuclear spins of constituent atoms are therefore decoupled from the electronic spin \cite{Mol-hyperf}, making them very insensitive to magnetic field 
fluctuations similar to recently proposed alkaline-earth atoms \cite{Alkaline-earths}.
Hyperfine or Zeeman sublevels of the ground rovibrational state are thus an especially attractive choice for qubit encoding. However, 
readout and initialization are expected to be more difficult for molecules
than for atoms. Due to the complex molecular level structure and the lack of closed (``cycling") transitions,  
resonant fluorescence used for readout in atoms is not in general applicable to molecules. Conditional conversion of a molecule into a 
pair of atoms (limited here to alkali dimers) followed by atomic detection
represents an attractive solution to this problem. In fact, coherent conversion back into atoms followed by atomic imaging has been used recently 
to detect ground state KRb molecules \cite{Ni}.
For rotational qubits conditional conversion can be realized by selectively exciting 
a specific rotational state with a laser field, since rotational splittings 
are typically in the GHz range and the states are well-resolved. For spin qubits one can first map the qubit state onto rotational sublevels and then 
apply the readout sequence.

We now discuss details of the scheme focusing on the implementation of the conditional atom-molecule conversion. 
The molecular state has to be stable to avoid decoherence, and deeply bound since the dipole moment reduces for high vibrational states. 
In the following we assume that atoms are transferred to the ground rovibrational molecular state.  
It is in general difficult to transfer atoms from a delocalized scattering state, even confined by a lattice, to a tightly localized 
deeply bound molecular state. Namely, the wavefunction overlap of the intermediate molecular state $\ket{v_{e}}$ with either the initial atomic or the 
final molecular state $\ket{g}$ will be small, requiring large intensity of the laser pulse to provide Rabi frequency sufficient for fast transfer. The 
problem of unfavorable wavefunction overlap can be alleviated by using Feshbach Optimized Photoassociation (FOPA) \cite{FOPA}, {\it i.e.} admixing 
a bound molecular state to the scattering atomic state, which can be realized 
by photoassociating atoms near a Feshbach resonance. For this, an external magnetic field can be tuned close to a 
Feshbach resonance for a specific qubit state. In this case the resonance 
strongly mixes the scattering state of an energetically open collision channel and a bound vibrational state of a closed channel 
as is illustrated in \fig{fig:STIRAP}. Admixure of the localized bound state to the motional state will strongly enhance the dipole moment for the 
transition to an excited molecular state $\ket{v_{e}}$ \cite{Our-NJP}. The required intensity of the pump field will be significantly reduced, 
making the transfer more efficient.  
 In the $^{87}$Rb+$^{7}$Li system, 
the wide resonance at 649 G in the $\ket{1,1}^{\rm Rb}+\ket{1,1}^{\rm Li}$ collision channel, recently observed in \cite{RbLi-Feshbach}, can be used. 
For this, Li has to be transferred to the ``enabled" $\ket{1,1}$ state, the conditional conversion will then take place from the $\ket{0'}$ qubit state.

During the atom-molecule conversion by an optical Raman $\pi$ pulse, the population in the molecular state is given by
$|c_{g}(t)|^{2}=\left(|\Omega_{R}|^{2}/(|\Omega_{R}|^{2}+\delta^{2})\right)\sin^{2}\left(\sqrt{|\Omega_{R}|^{2}+\delta^{2}}t/2\right)$,
where $\Omega_{R}=\Omega_{p}\Omega_{S}/2\Delta_{e}$ is the Rabi frequency of the pulse, $\Omega_{p,S}$ are the Rabi frequencies of the pump and Stokes 
fields, and $\Delta_{e}$ is the common one-photon detuning of the fields from the excited molecular state.
If there is an external electric field to orient the dipole moments of molecules in each site, they will interact via 
dipole-dipole interaction, leading to an accumulated phase $\phi(t)=\int_{0}^{t}V_{\rm dd}|c_{g}(t')|^{4}dt'/\hbar$ of the $\ket{g,g}$ 
state (the other combinations do not acquire a phase, since there is no electric dipole moment in at least one of the states). Here 
$V_{\rm dd}=\mu_{\rm ind}^{2}/r^{3}$ 
is the dipole-dipole interaction strength, $\mu_{\rm ind}\approx \mu\left(\mu E_{\rm dc}/3\hbar B\right)$ is the dipole moment induced 
in the ground rovibrational state by an electric field $E_{\rm dc}$, $\mu $ is the permanent dipole moment, $\hbar B$ is the rotational energy of the molecular state, and $r$ is the 
distance between two molecules. After a 
specific time we can reverse the process and convert the molecules back into atoms $\ket{g}\rightarrow \ket{0'}$ such that the total accumulated phase 
$\phi=\pi$, 
and finally rotate the ``enabler" atoms $^{7}$Li back to their ``inert" state $\ket{2,2}$. This sequence of operations leads to a phase gate.
We can estimate the duration of one and two-qubit gates in our system. 
Eigenstates and eigenenergies of a system of two different atoms in a deep lattice site interacting via a short range van del Waals 
interaction have been calculated in \cite{Ospelkaus-rf}. 
The eigenenergies and eigenstates depend on the interaction strength via a background {\it s}-wave scattering length $a_{\rm bg}$. 
Since $a_{\rm g}$ will in general be different for the $\ket{0}$ and $\ket{1}$ qubit states, 
the motional states will differ as well. As a result to avoid undesirable entanglement between internal and motional states 
one-qubit rotations have to be performed 
adiabatically compared to the oscillation period of the lattice, which limits the one-qubit rotation time by tens $\mu$s for traps with 
$\sim 100$ kHz oscillation frequency. 
The duration of the optical Raman $\pi$ pulse transferring the Li atom to the ``enabled" $\ket{1,1}$ state is 
limited by the same requirement. The Raman pulse converting atoms to molecules has to be adiabatic as well to avoid coupling of internal and motional states by the 
dipole-dipole interaction, leading to an error in the accumulated phase. Additionally, its Rabi frequency has to be much larger than the shift of the 
molecular state due to dipole-dipole interaction for the $\pi$ pulse to be in a two-photon resonance. The phase is accumulated during two 
Raman $\pi$ pulses and interaction time $\tau_{\rm int}$, with the total phase given by $\phi=V_{\rm dd}(3\pi/4\sqrt{|\Omega_{R}|^{2}+\delta^{2}}+\tau_{\rm int})/\hbar$. 
Assuming that the pump and Stokes fields are in a two-photon resonance when there is no dipole-dipole interaction, 
so that $\delta=V_{\rm dd}/\hbar$, and $|\Omega_{R}|>>\delta$, the interaction time required to accumulate the $\pi$ phase 
shift is $\tau_{\rm int}=\left(\pi \hbar/V_{\rm dd}\right)\left(1-3V_{\rm dd}/4|\Omega_{R}|\right)\approx \pi\hbar/V_{\rm dd}$.   
For two LiRb molecules with induced dipole moment of the order of the permanent one $\mu =4.2$ D \cite{LiRb-dip-mom}, formed in neighboring lattice sites 
separated by $r=\lambda/2=500$ nm, the dipole-dipole 
interaction strength is $V_{\rm dd}=\mu ^{2}/r^{3}\sim 10^{5}$ s$^{-1}$. Choosing the $\pi$ pulse Rabi frequency $\Omega_{R}=10^{6}$ s$^{-1}$ ($\gg V_{\rm dd}$) 
the resulting 
pulse duration $\sim 3$ $\mu$s, and the $\pi$ phase accumulation time $\tau_{\rm int}=\pi\hbar/V_{\rm dd}\sim 14$ $\mu$s, giving the total phase gate time $\sim 20$ $\mu$s. 
%Provided that the dipole moment ratio $\eta=|\bra{v_{e}}\vec{\mu}\ket{\omega^{0}_{op}}|/|\bra{v_{e}}\vec{\mu}\ket{v_{cl}}|\sim 10^{-2}-10^{-3}$ the 
%populaion in the $\ket{g,g}$ state is $|c_{fin}|^{2} \sim \eta^{2}\sim 10^{-4}-10^{-6}$, which gives the phase error during the gate.  
Molecule-atom conversion for molecular 
qubit readout is limited by the requirement of state selectivity, i.e. the bandwidth and the Rabi frequency of the laser field 
must be much smaller than the qubit transition frequency. Hyperfine splittings in the ground rovibrational state of $^{1}\Sigma^{+}$ potential are 
expected to be in the kHz range \cite{Mol-hyperf}, resulting in readout pulses of ms duration. These can be shortened to ns by mapping the spin qubit states 
onto rotational sublevels, having splittings in GHz range.   
We note that a trapping potential experienced by a pair of atoms and a deeply bound molecule will in general differ. Non-adiabatic conversion of atoms into a molecule 
will then result in excitation of a number of motional states in the molecular trapping potential and oscillations of the molecular center-of-mass 
wavepacket \cite{STIRAP-oscill}. To prevent the oscillations, which will lead to uncertainty in the accumulated phase, molecules can be 
trapped in a separate potential, not affecting atoms, and vice versa.

One and two qubit gates in our scheme require single-site addressing, which can be realized using a magnetic field gradient. 
When an external magnetic field is applied the energies of the qubit states are given by the Breit-Rabi formula
\begin{equation}
\frac{E(f=I\pm 1/2,m)}{\Delta E_{\rm hf}}=-\frac{1}{12}\pm\frac{1}{2}\sqrt{1+mx+x^{2}},
\end{equation}
where $x=g_{J}\mu_{B}B/\Delta E_{\rm hf}$, and for $^{87}$Rb $\Delta E_{\rm hf}=6.835$ GHz. The qubit transition frequency 
at 649 G corresponding to the Feshbach resonance value is $\Delta E=E(2,2)-E(1,1)=8.3$ GHz, and varies with the magnetic field as $\approx 2.38$ MHz/G. 
With magnetic field gradients of several kG/cm, realizable experimentally, 
the change of the qubit frequency between two sites separated by 500 nm will be $\sim 100$ kHz, and neighboring sites can be frequency resolved. 
At the same time using a wide Feshbach resonance of several G width the resonance condition will be satisfied for $\sim 10^{2}$ lattice sites in each 
lattice dimension.

Finally, we discuss major decoherence mechanisms of the proposed scheme. Decoherence of qubit states induced by the optical lattice (elastic and inelastic scattering 
of lattice photons, ac Stark shifts of hyperfine levels, laser intensity and frequency fluctuations) can be reduced to sub-Hz \cite{Saffman-Rb}. 
For hyperfine states magnetic field fluctuations might affect the 
qubit coherence. Currently fluctuations of magnetic field 
can be suppressed down to 300 $\mu$G level, which will result in the dephasing time of the qubit transition of $\approx 200$ $\mu$s. 
With two-qubit gate times 
$\sim 20$ $\mu$s, $n\approx 10$ operations can be realized. To reduce the qubit sensitivity to magnetic field fluctuations, another combinations 
of alkali atoms can be chosen. For example, encoding a qubit into states $\ket{f=I\pm 1/2,m=-f}$ with most negative spin projections will 
significantly reduce the dependence of the qubit frequency on the magnetic field.   
To reduce the sensitivity even further, alkaline-earth atoms could be used, having $^{1}S_{0}$ and 
$^{3}P_{0}$ optical clock states with vanishing coupling beween electronic and nuclear spins, and the latter can be utilized for qubit encoding 
\cite{Alkaline-earths}. Since alkaline-earth atoms do not have magnetic Feshbach resonances, optical Feshbach resonances can be used instead to enhance the 
phase gate, although the possibility to associate two different alkaline-earth atoms into a molecule has not yet been explored. 
Another source of decoherence is 
inelastic collisions between atoms. Storage qubit states have to be chosen such that collisions are elastic in these states. In the Rb-Li system 
the qubit states $\ket{2,2}^{\rm Rb}+\ket{2,2}^{\rm Li}$ and $\ket{1,1}^{\rm Rb}+\ket{2,2}^{\rm Li}$ 
are stable to collisions since the total $m_{\rm tot}=m_{\rm Rb}+m_{\rm Li}$ is conserved and there are no other energetically open channels with 
$m=4$ and $3$ at ultralow temperatures, respectively. Inelastic decay rate in the ``enabled" 
qubit states has to be slow compared to the two-qubit gate duration to minimize the gate error. In Rb-Li the $\ket{1,1}^{\rm Rb}+\ket{1,1}^{\rm Li}$ channel 
is elastic because other channels with $m_{\rm tot}=2$ are closed, but the $\ket{2,2}^{\rm Rb}+\ket{1,1}^{\rm Li}$ channel can inelastically 
decay to the $\ket{1,1}^{\rm Rb}+\ket{2,2}^{\rm Li}$ state. The inelastic 
collisional rate for this channel is $\sim 10^{5}$ s$^{-1}$ close to the Feshbach resonance, resulting in high error during the phase gate of 20 $\mu$s
duration. We note that the inelastic decay rate varies among alkali atoms, and other systems might have smaller rates.

To conclude, we present a novel hybrid atomic/molecular system for quantum computation, which combines the principal advantages of neutral atom and 
polar molecule-based approaches. Encoding a qubit in atomic states allows easy initializion, readout, one-qubit 
operations, as well as mapping the qubit state onto a photon for quantum communication. On the other hand, conditional conversion of a pair of atoms into a 
polar molecule leads to strong dipole-dipole interaction, resulting in fast two-qubit gates. The conditional conversion can also 
be used for non-destructive reading out of molecular qubit states. It is worth noting that one of the advantages 
of the proposed setup is that all elements are realizable with existing technology.  
 
The authors gratfully acknowledge fruitful discussions with M. Lukin and financial support from ARO and NSF.

\end{document}